\documentclass{article}

\usepackage{arxiv}

\usepackage[utf8]{inputenc} 
\usepackage[T1]{fontenc}    
\usepackage{hyperref}       
\usepackage{url}            
\usepackage{booktabs}       
\usepackage{amsfonts}       
\usepackage{nicefrac}       
\usepackage{microtype}      
\usepackage{lipsum}
\usepackage{amsmath}
\usepackage[T2A,T1]{fontenc}
\usepackage[utf8]{inputenc}
\usepackage[russian,english]{babel}
\usepackage{cases}
\usepackage{graphicx}
\usepackage{wrapfig}
\usepackage{bm}

\title{Translation: About one scheme of tomography}

\begin{document}
\maketitle

\begin{center}
\foreignlanguage{russian}{ИЗВЕСТИЯ ВЫСШИХ УЧЕБНЫХ ЗАВЕДЕНИЙ -- РАДИОФИЗИКА}\\
\foreignlanguage{russian}{Том 1, № 3 -- 1958}\\
\foreignlanguage{russian}{ОБ ОДНОЙ СХЕМЕ ТОМОГРАФИИ}\\
\foreignlanguage{russian}{Б. И. Коренблюм, С. И. Тетельбаум, А. А. Тютин}\\
\hspace{1cm}
\hrule

PROCEEDINGS OF HIGHER EDUCATIONAL INSTITUTIONS - RADIOPHYSICS

Volume 1, No. 3 -- 1958

ABOUT ONE SCHEME OF TOMOGRAPHY

B. I. Korenblum, S. I. Tetelbaum, A. A. Tyutin

\hspace{1cm}

Translation:
Alex Gustschin\footnote{Chair of Biomedical Physics, Department of Physics and Munich School of Bioengineering, Technical University 
of Munich, 85748, Garching, Germany, alex.gustschin@ph.tum.de}

\hspace{1cm}

\hrule
\end{center}

\begin{abstract}
A new method to acquire cross-sectional x-ray images is studied which is based on processing information contained in an x-ray sinogram\footnote{Original: \foreignlanguage{russian}{рентгеношифрограмм}, „encoded x-ray radiograph”. This term refers to what is now called x-ray sinogram. Although the term “sinogram” does not appear in the original work in adequate Russian linguistic morphology, it is clear that projectional radiographs from an examined plane at continuous sample rotation is referred to. In the following \foreignlanguage{russian}{рентгеношифрограмм} will be translated as (x-ray) sinogram.} recorded at varying angles of an object. A derivation of the respective integral equation and its solution are given as well as a functional scheme of a computing and television screen-based device enabling to acquire a selected cross-sectional x-ray images of the object.

\end{abstract}

\section*{Introduction}

A characteristic feature of tomography methods used so far is that the shadow image of an examined layer in an object is superimposed with the images of its other layers, “blurred” to a greater extent the farther these layers are apart from the studied layer. This issue significantly limits the possibilities of x-ray-based examination of sufficiently thin layers and obtaining sharp layer-by-layer images of the object. Nevertheless, tomography and tomofluorography are used in several important practical applications  \cite{ref1}.

Based on envisioned possibilities to compensate image distortion \cite{ref2}, methods of x-ray technology were proposed which allow the determination of the local x-ray attenuation coefficient in each element of a three-dimensional object and to obtain a volumetric image of the latter \cite{ref3, ref4}. Employing these methods, radiographic examination schemes of practically any desired layer thickness can also be realized, and the image acquired will not depend on other characteristic features of other layers in the object.

The present article studies one such scheme. The most important practical case of a relatively close radiation source resulting in an object irradiated by a narrow fan-shaped beam is studied. During a continuous rotation of the object around an axis perpendicular to the layer under investigation, a photosensitive film moving parallel to the axis acquires encoded x-ray radiograms which contain the necessary data to obtain images of the layer. A television screen-equipped analog computing device implemented to solve the respective integral equation converts the information contained in the x-ray sinogram into an image of the layer in the object under study. 

The article derives the integral equation of the problem and gives its solution. Further, a block diagram of the respective computing device is described.

\section{Acquisition of sinogram and integral equation}

\begin{figure}
  \centering
  \includegraphics[width=13cm]{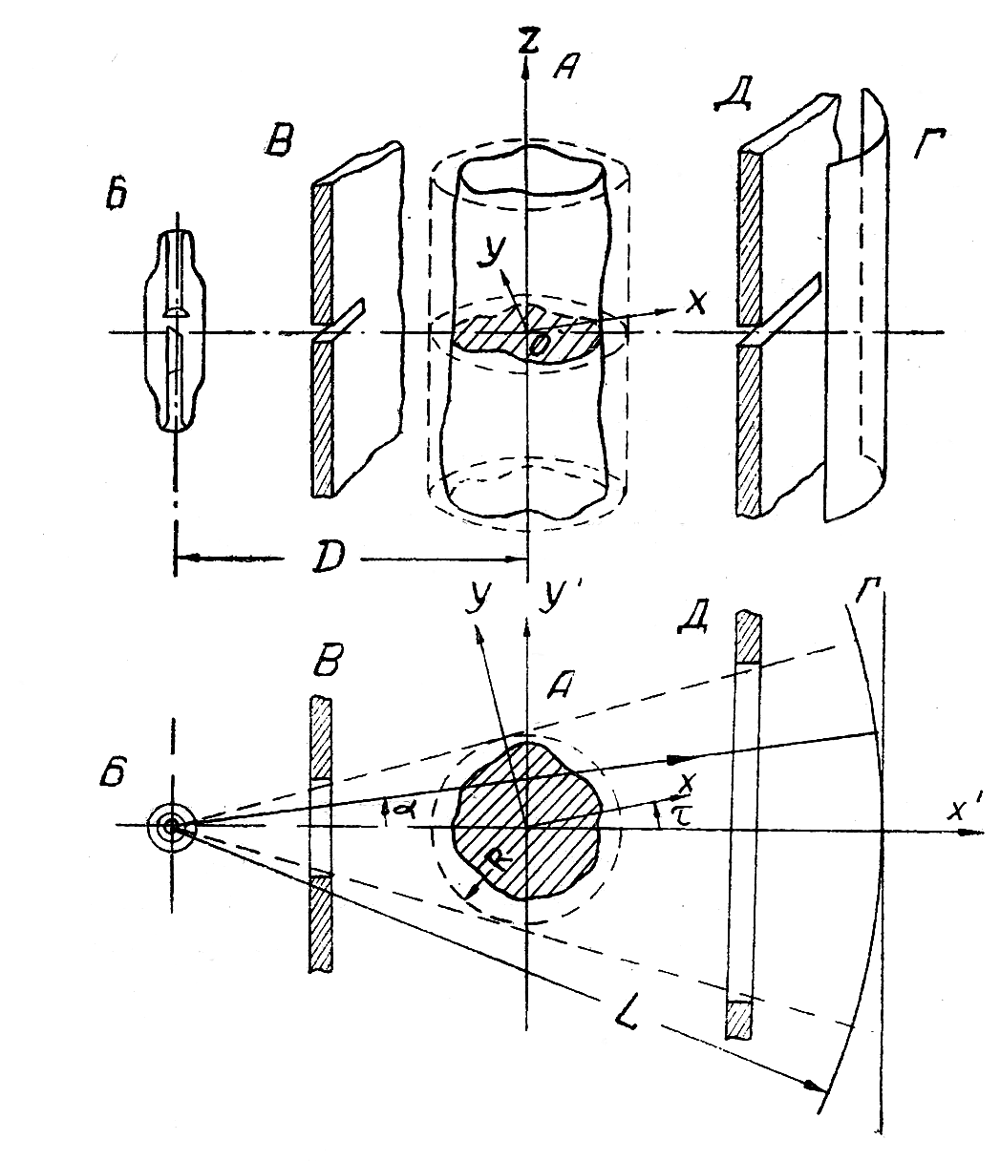}
  \caption{Schematic setup of the apparatus and the object to obtain the x-ray sinogram. }
  \label{fig:fig1}
\end{figure}

Let object A (Fig. \ref{fig:fig1}) be irradiated with a sufficiently narrow fan-shaped x-ray beam originating in a monochromatic source \foreignlanguage{russian}{Б} (in plane of the investigated section) and shaped by a slot collimator \foreignlanguage{russian}{В}\footnote{In Fig.\ref{fig:fig1} a second slot collimator \foreignlanguage{russian}{Д} is implemented to shield the film from scattered x-ray radiation.}. To obtain the x-ray image of the layer it is necessary to determine the values of the locally distributed x-ray attenuation coefficient $F(x, y)$, which is a function of the geometric coordinates of the object, located in the $xy$ plane (see Fig. \ref{fig:fig1}).

We add an object enclosed by a cylinder of radius $R$ and an axis $z$ perpendicular to the $xy$ plane and set $F(x,y)$ inside the cylinder but outside the object equal to zero. 
We shall determine the radiation intensity $F(\alpha, \tau)$ transmitted by the object and illuminating the x-ray film \foreignlanguage{russian}{Г} which is arranged on a cylindrical surface of radius $L$ with an axis parallel to $z$-direction through the source \foreignlanguage{russian}{Б}. The intensity to be determined will be inversely proportional to $L^2$ and decrease with increasing attenuation of the radiation caused by all elements of the object located on the linear path of each ray. It can be expressed up to a constant factor $K$:

\begin{equation} \label{eq:1}
F(\alpha, \tau) = \frac{K}{L^2} \exp \left[ -\sqrt{1+\tan^2\alpha} \int\limits_{x'_1}^{x'_2} F(x,y) \ dx' \right], 
\end{equation}

where 
\begin{equation*}
\begin{aligned}
 x = x' \cos\tau + (D+x') \tan \alpha \sin \tau ;
 \end{aligned}
\end{equation*}

\begin{equation*}
\begin{aligned}
  y = -x' \sin \tau + (D+x') \tan \alpha \cos \tau ;
 \end{aligned}
\end{equation*}

\begin{equation*}
\begin{aligned}
  x'_{1,2} = \frac{-D\tan^2 \alpha \pm \sqrt{R^2 + (R^2-D^2) \tan^2\alpha}}{1 + \tan^2\alpha}. 
 \end{aligned}
\end{equation*}

After the exposure of the film, it will contain an image in the shape of a narrow stripe which characterizes the function $F(\alpha, \tau)$ for a fixed $\tau$.
We will rotate the object with constant angular velocity around the $z$-axis and simultaneously move the film at a velocity $v$ along the $z$-axis. In this case, the function $F(\alpha, \tau)$ will be recorded on the film with a smoothly alternating angle $\tau$ – an x-ray sinogram containing necessary and sufficient information to obtain the layer's x-ray image. The dependence of the film’s darkening density on exposure can also be taken into account.

The possibility of obtaining an x-ray image of the layer, i.e. the determination of the unknown function $F(x, y)$ (characterizing the distribution of the local x-ray attenuation coefficient in the cross-section of the object) on the basis of the function $F(\alpha, \tau)$ found from experiment is a result of the uniqueness of the continuous solution of the integral equation.

\section{Solution of the integral equation }

We assume that the function $F(x,y)$ is continuously differentiable three times. After taking the logarithm, Equation (\ref{eq:1}) can be expressed as:

\begin{equation} \label{eq:2}
\begin{aligned}
  \int\limits_{l}^{} F(x,y) \ dl =  \Phi (\eta, \sigma)
 \end{aligned}
\end{equation}

where integration is carried out along line $l$, which is parameterized by

\begin{equation*}
 \begin{aligned}
  x \sin \sigma + y \cos\sigma = \eta,
 \end{aligned}
\end{equation*}

moreover

\begin{equation} \label{eq:3}
\begin{aligned}
  \sigma = \tau - \alpha ; \eta = D \sin\alpha;
 \end{aligned}
\end{equation}

\begin{numcases}{\Phi(\eta, \sigma)=}-\ln F(\alpha, \tau) +\ln \frac{K}{L^2}  & for $(\mid\eta\mid\ \leq R)$ \nonumber \\ 0 & for $(\mid\eta\mid  > R)$
\label{eq:4}
\end{numcases}

So, our task comes down to the following: knowing the integral of the Function $F(x,y)$ along all possible lines $l$, find the function $F(x,y)$ itself. 
Similar problems were studied in several works (see e.g. \cite{ref5, ref6}). Below, a solution of the problem is given based on the theory of the Fourier integral. Let us find the Fourier transform of the unknown function $F(x, y)$:     

\begin{equation}  \label{eq:5}
\begin{aligned}
  g(u,v) = \iint \limits_{-\infty}^{+\infty} F(x,y)e^{-i\bm{r_1 r}} \ dx dy
 \end{aligned}
\end{equation}

where $ \bm{r_1} = (u,v); \bm{r} = (x,y).$ Integral (\ref{eq:5}) can be converted as follows:

\begin{equation} \label{eq:6}
\begin{aligned}
  g(u,v) = \frac{1}{|\bm{r_1|}} \int \limits_{-\infty}^{+\infty} e^{-it} \ dt \int \limits_{\bm{r_1 r} =t}^{} F(x,y) \ dl
 \end{aligned}
\end{equation}

Assuming

\begin{equation} \label{eq:7}
\begin{aligned}
  u = \rho \cos{\theta}; v = \rho \sin{\theta} 
  \end{aligned}
\end{equation}

and using (\ref{eq:2}), (\ref{eq:4}) we rewrite (\ref{eq:6}) as follows:

\begin{equation} \label{eq:8}
\begin{aligned}
  g(u,v) = \frac{1}{\rho} \int \limits_{-\infty}^{+\infty} e^{-it} \Phi\Bigl(\frac{t}{\rho}, \frac{\pi}{2} - \theta \Bigr) \ dt =  \int \limits_{-\infty}^{+\infty} e^{-i \rho \eta} \Phi\Bigl(\eta, \frac{\pi}{2} - \theta \Bigr) \ d\eta.
 \end{aligned}
\end{equation}

As known from the Fourier integral theory, $g(u, v)$ decreases at least with $\rho^{-3}$ for $\rho \rightarrow \infty$. Therefore, we can apply the inversion formula of the Fourier integral

\begin{equation} \label{eq:9}
\begin{aligned}
  F(x,y) = - \frac{1}{4 \pi^2} \iint \limits_{-\infty}^{+\infty} e^{i \bm{r_1 r}} \Bigl\lbrack \int \limits_{-\infty}^{+\infty} e^{-i\rho \eta} \Phi\Bigl(\eta, \frac{\pi}{2} - \theta \Bigr)  \ d\eta \Bigr\rbrack \ du dv,
 \end{aligned}
\end{equation}

where the outer double integral converges absolutely. However, it is not possible to change the order of integration here, because the triple integral is not absolutely convergent. Therefore, we introduce the convergence factor $e^{-\delta|\bm{r_1}|}$ and move to the limit for $\delta \rightarrow 0$. Changing to polar coordinates and denoting

\begin{equation} \label{eq:10}
\begin{aligned}
 x = r \cos \varphi; y = r \sin \varphi,
 \end{aligned}
\end{equation}

we get from (\ref{eq:9}):

\begin{equation} \label{eq:11}
\begin{aligned}
 F(x,y) = - \frac{1}{4 \pi^2} \lim_{\delta \to + 0} \int  \limits_{0}^{2\pi} \int  \limits_{0}^{+\infty} \Bigl\lbrack  \int \limits_{-\infty}^{+\infty} e^{-i \rho \eta} \Phi\Bigl(\eta, \frac{\pi}{2} - \theta \Bigr) \ d\eta \Bigr\rbrack \ e^{-\delta \rho} e^{ir \cos (\theta-\varphi)} \rho \ d\rho d\theta \\\
  = - \frac{1}{4 \pi^2} \lim_{\delta \to + 0} \int  \limits_{0}^{2\pi} \int  \limits_{-\infty}^{+\infty}
 \frac{\Phi\Bigl(\eta, \frac{\pi}{2} - \theta \Bigr) \ d\eta d\theta}{\lbrack r \cos{(\theta -\varphi)} - \eta + \delta i \rbrack^2}
 \end{aligned}
\end{equation}

Since the functions $F(x, y)$ and $\Phi(\eta, \sigma)$ are real-valued, the imaginary part in the integrative expression can be discarded (\ref{eq:11}):

\begin{equation} \label{eq:12}
\begin{aligned}
 F(x,y) = - \frac{1}{4 \pi^2} \lim_{\delta \to + 0} \int  \limits_{0}^{2\pi} \int  \limits_{-\infty}^{+\infty}
 \frac{\lbrack r \cos{(\theta -\varphi)} - \eta \rbrack^2 - \delta^2 }{\{\lbrack r \cos{(\theta -\varphi)} - \eta \rbrack^2 + \delta^2\}^2}  \Phi\Bigl(\eta, \frac{\pi}{2} - \theta \Bigr)\ d\eta d\theta \\\
  = - \frac{1}{4 \pi^2} \lim_{\delta \to + 0} \int  \limits_{0}^{2\pi} \int  \limits_{-\infty}^{+\infty}
 \frac{\lbrack r \sin{(\varphi +\sigma)} - \eta \rbrack^2 - \delta^2 }{\{\lbrack r \sin{(\varphi +\sigma)} - \eta \rbrack^2 + \delta^2\}^2}  \Phi\Bigl(\eta, \sigma\Bigr)\ d\eta d\sigma
 \end{aligned}
\end{equation}

To calculate the limit here, we will use the following lemma.

\textbf{Lemma.} If $f(x)$ is twice differentiable and bounded on $(-\infty,+\infty)$, and if $f'(x)$ decreases faster for $ |x| \rightarrow \infty$ than some degree of $|x|^{-\alpha} (\alpha>0)$, then:

\begin{equation*}
\begin{aligned}
 \lim_{\delta \to + 0} \int  \limits_{-\infty}^{+\infty} \frac{x^2+\delta^2}{(x^2+\delta^2)^2} f(x) \ dx =  \int  \limits_{-\infty}^{+\infty} \frac{f'(x)}{x} \ dx,
 \end{aligned}
\end{equation*}

where the integral on the right is understood in the sense of the Cauchy principal value:

\begin{equation*}
\begin{aligned}
\int  \limits_{-\infty}^{+\infty} \frac{f'(x)}{x} \ dx =  \lim_{\epsilon \to + 0} 
\biggl\lbrack \biggl( \int  \limits_{-\infty}^{-\epsilon} + \int  \limits_{\epsilon}^{+\infty} \biggr ) \frac{f'(x)}{x} \ dx  \biggr\rbrack = \int  \limits_{0}^{+\infty} \frac{f'(x)-f'(-x)}{x} \ dx
 \end{aligned}
\end{equation*}

The proof of this lemma is omitted. 

Applying this lemma, we transform (\ref{eq:12}) to the form
\begin{equation} \label{eq:13}
\begin{aligned}
F(x,y) = -\frac{1}{4 \pi^2} \int  \limits_{0}^{2\pi} \ d\sigma \int  \limits_{-\infty}^{+\infty} \frac{\frac{\partial}{\partial\eta} \Phi(\eta,\sigma) \ d\eta} {\eta - r \sin{(\varphi + \sigma)}},
 \end{aligned}
\end{equation}

where the integral is considered in context of the principal value relative to the point $\eta =r \sin(\varphi+\sigma)$. For example, let's assume 

\begin{numcases}{\Phi(\eta, \sigma)=}1  & for $(\mid\eta\mid\ < R)$ \nonumber \\ 0 & for $(\mid\eta\mid > R)$. \nonumber 
\end{numcases}

In this case the derivative $\frac{\partial}{\partial\eta} \Phi(\eta, \sigma) = \delta(\eta + R) - \delta(\eta-R)$ is the difference between two Dirac functions. Equation (\ref{eq:13}) gives

\begin{equation*}
\begin{aligned}
F(x,y) = \frac{1}{4 \pi^2} \int \limits_{0}^{2\pi} \biggl\lbrack \frac{1}{R-r\sin(\varphi + \sigma) }+ \frac{1}{R+r \sin{(\varphi +\sigma)}} \biggr\rbrack  \ d\sigma = \frac{1}{\pi \sqrt{R^2-r^2}} \\\
(|r|<R).
 \end{aligned}
\end{equation*}

Note that for $|\bm{r}| > R$ formula (\ref{eq:13}) cannot be used directly, because for  values of $\sigma$ at which $r \sin(\varphi+\sigma) = \pm R$, the integral

\begin{equation*}
\begin{aligned}
\int \limits_{-\infty}^{+\infty} \frac{\frac{\partial}{\partial\eta} \Phi(\eta,\sigma) } {\eta - r \sin{(\varphi + \sigma)}} \ d\eta 
 \end{aligned}
\end{equation*}

is invalid. However, formula (\ref{eq:12}) also applies at $|\bm{r}| > R$. In general formula (\ref{eq:12}) does not assume continuity, let alone differentiability of the function $\Phi(\eta, \sigma)$.

Now the solution of equation (\ref{eq:1}) is simple. For that $\Phi(\eta, \sigma)$ from function (\ref{eq:4}) is inserted in (\ref{eq:13}) and the expression is transformed to variables $\alpha$ and $\sigma$ by formula (\ref{eq:3}). We obtain

\begin{equation} \label{eq:14}
\begin{aligned}
F(x,y) = \frac{1}{4 \pi^2} \int \limits_{0}^{2\pi}  \ d\tau \int \limits_{-\arcsin{(R/D)}}^{+\arcsin{(R/D)}}  \frac{(\frac{\partial}{\partial\alpha} + \frac{\partial}{\partial\tau}) \ln F(\alpha, \tau)}{(x' + D) \sin \alpha - y'\cos \alpha }\ d\alpha 
 \end{aligned}
\end{equation}

or after changing variables $\alpha, \tau$ to $\alpha, \alpha+\sigma$:

\begin{equation} \label{eq:15}
\begin{aligned}
F(x,y) = \frac{1}{4 \pi^2} \int \limits_{0}^{2\pi}  \ d\sigma \int \limits_{-\arcsin{(R/D)}}^{+\arcsin{(R/D)}}  \frac{\frac{\partial}{\partial\alpha} \ln F(\alpha, \alpha+\sigma)}{D \sin\alpha - (x\sin \sigma + y \cos\sigma)} \ d\alpha 
 \end{aligned}
\end{equation}

where

\begin{equation*}
 \begin{aligned}
  x'= x \cos\tau -y \sin \tau,
 \end{aligned}
\end{equation*}

\begin{equation*}
 \begin{aligned}
  y'= x \sin\tau + y \cos \tau,
 \end{aligned}
\end{equation*}

(Inner integrals in (\ref{eq:14}) and (\ref{eq:15}) are considered in context of the Cauchy principal value relative to the points $\alpha = \arctan \frac{y'}{x'+D}$ and $\alpha = \arcsin{[\frac{1}{D}(x \sin{\sigma} + y\cos{\sigma})]}$ respectively.) In this case, the function $F(\alpha, \tau)$ should be considered to be periodically continued by the second argument outside the main interval $0 \leq \tau \leq 2\pi$.

\section{Functional block diagram of television screen-equipped computing device}

\begin{figure}
  \centering
  \includegraphics[width=10cm]{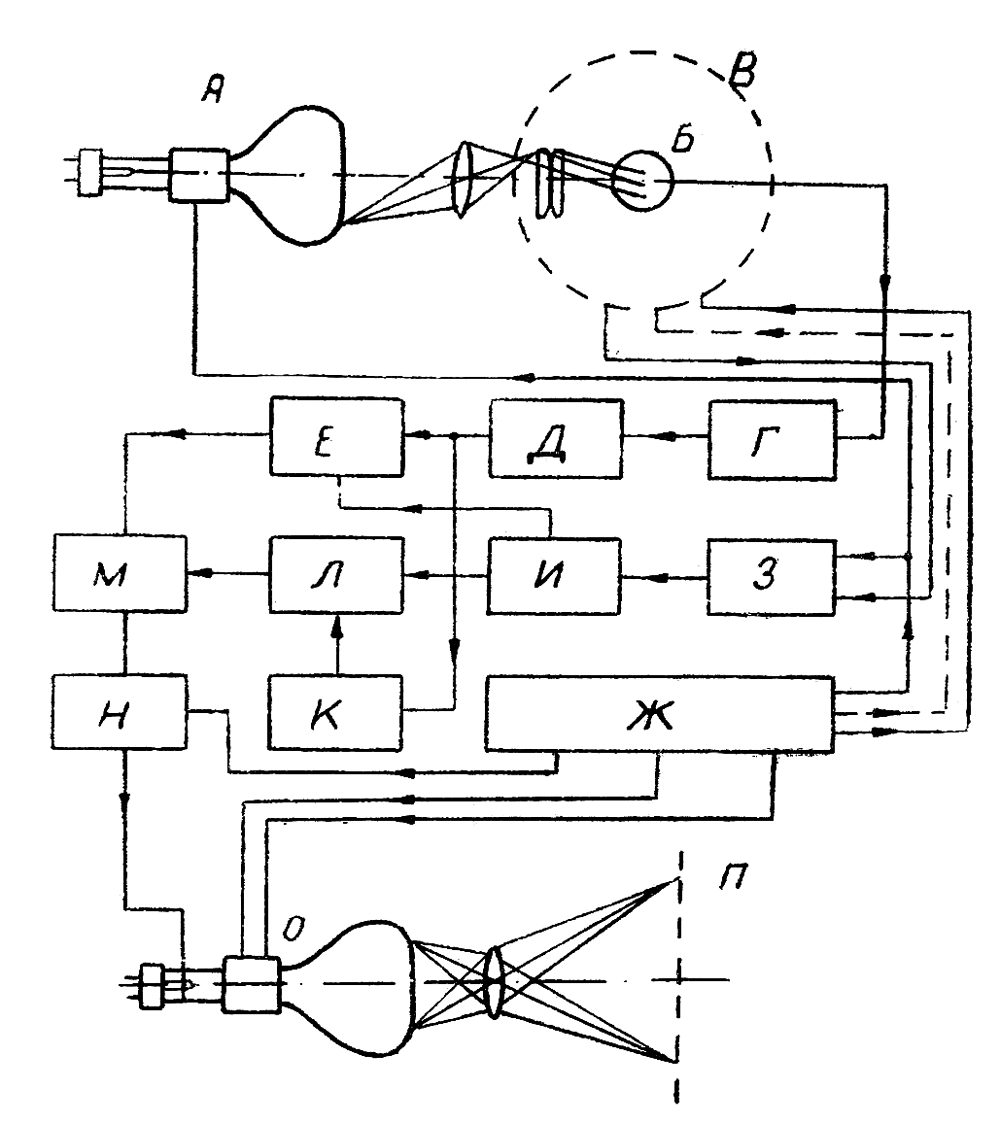}
  \caption{Functional block diagram of the television-based computing device.}
  \label{fig:fig2}
\end{figure}

In the design of the block diagram (Fig. \ref{fig:fig2}) it was taken into account that the integrand function has a simple pole within the integration interval of the variable $\alpha$.

The x-ray sinogram is "read" in a sensor consisting of a transferring television tube (A), a photo-multiplier (\foreignlanguage{russian}{Б}) and a mechanical system for the movement of the sinogram film mounted on a rotating drum (\foreignlanguage{russian}{В}). The sensor output is an electric signal (video signal) whose amplitude at time $t$ corresponds to the density of blackening of the sinogram $F(\alpha, \alpha+\sigma)$ at the point with coordinates $(\alpha , \sigma)$ and goes to the logarithmizing cascade (\foreignlanguage{russian}{Г}), then it is differentiated (cascade \foreignlanguage{russian}{Д}) and conducted to the multiplier E. Simultaneously, the sweep block (\foreignlanguage{russian}{Ж}), which is coupled to the potentiometric sensors (providing voltages proportional to $\sin \sigma$ and $\cos \sigma$) of the mechanical rotation system of the drum \foreignlanguage{russian}{В}, guides the signal though the summation cascade (\foreignlanguage{russian}{З}) and the generator of the inverse function (\foreignlanguage{russian}{И}), supplying the multiplier with a voltage corresponding to the inverse denominator of the integrand.

To model a function with a pole, the following method is used. The integral with the integration limits located on both sides of the pole ($\alpha_i \pm \epsilon$) is extracted and replaced by an approximation:

\begin{equation*}
\begin{aligned}
F(x,y) = \int \limits_{\alpha_i - \epsilon}^{\alpha_i + \epsilon}  \frac{\frac{\partial}{\partial\alpha} \ln F(\alpha, \alpha+\sigma)}{D \sin\alpha - (x\sin \sigma + y \cos\sigma)} \ d\alpha  \approx 2 \epsilon \Bigl| \frac{\partial^2}{\partial \alpha^2} \ln F(\alpha, \alpha+\sigma) \Bigl|_{\alpha=\alpha_i}
 \end{aligned}
\end{equation*}

where
\begin{equation*}
\begin{aligned}
\alpha_i = \arcsin{\Bigl[\frac{1}{D}(x \sin{\sigma} + y\cos{\sigma})\Bigr].}
 \end{aligned}
\end{equation*}

The second differentiating stage (K) and the "switch" (\foreignlanguage{russian}{Л}) are used to obtain this term, "cutting out" the part corresponding to the interval $2\epsilon$ from the output voltage of the differentiating cascade. The switch is controlled by the generator of the inverse function, so that the interval of $2\epsilon$ is provided automatically.

The multiplier output voltage (E) is added to the voltage from the switch output (\foreignlanguage{russian}{Л}), integrated by the integrator (M) and supplied to switch (H). The determination of the desired value of the function $F(x, y)$ for each element of the object consists of a single calculation cycle of the computing device. The switch (H) controlled by the sweep generator (\foreignlanguage{russian}{Ж}) sets the same initial conditions for each calculation cycle. From the switch (H) the voltage corresponding to the searched function $F(x, y)$ is supplied to the control grid of the television screen receiver, which is connected to the sweep generator system. 

As a result, a spot appears at the screen in a certain point with coordinates $(x, y)$ and its brightness is proportional to the x-ray attenuation coefficient $F(x, y)$ in the corresponding element of the object cross-section. The image from the television screen is projected on a photographic film (\foreignlanguage{russian}{П}). 

Logarithmic, differentiating and other cascades of the analog computation device are designed by known principles of mathematical modeling \cite{ref7}.

It should be noted that a significant amount of information must be processed to obtain an image of the layer. Therefore, at the speed of operation, which is adapted to the after-glow time of modern television tubes designed for scanning beam systems ($0.3 \mu s$) and acceptable rotation speeds of the drum (3000 rpm), the image of a layer with a resolution of $10^4$ elements can be obtained in about 5 minutes. For that the circuitry of the analog computing device should be designed for a frequency bandwidth of up to 1 MHz.

Considerations about the possibility of examination using the continuous spectrum of x-ray frequencies and the use of selective color contrasting are described in \cite{ref3, ref4}.

Currently, the first experimental system for obtaining x-ray images of thin layers according to the scheme described in this article is under construction at the Kiev Polytechnic Institute.

\begin{center}
    Kiev Polytechnic Institute\\
    Submitted January 7, 1958\\
\end{center}

\begin{center}
    -END OF ORIGINAL ARTICLE-
\end{center} 

\section{Translator's Note}
The early efforts in x-ray tomography at the Kiev Polytechnical Institute reported in the present translated work are remarkable in several aspects. The idea to reconstruct three-dimensional representations of the local x-ray attenuation coefficient from angular x-ray projections was first reported by Tetelbaum in 1957 (Ref. 3 in original work, translation available \cite{nref6}) and deduced in detail in the present article for the special case of a fan-beam geometry. It is noteworthy that Allan Cormack, who is considered to be the theoretical pioneer of computed tomography, published this idea and the solution to the problem in 1963 \cite{nref1} – more than 5 years after Tetelbaum and Korenblum. After some experiments with gamma sources, Cormack did not pursue further development due to the lack of interest in the community until hearing about Hounsfield’s work \cite{nref2}. It also seems that neither Cormack and Hounsfield nor Korenblum and coworkers were familiar with Radon’s work \cite{nref3} on respective integral transforms, which provided a mathematical solution to the inverse problem, however, did not envision any practical applications. 
It is evident from the present paper that Korenblum et al. did not only present the mathematical solution to the problem but were also sincerely working towards an experimental realization of such a device with technical equipment available in the 50s. In contrast to first CT scanner designs in the early 70s in the West ('rotate-translate scanners'), the design proposed here was a fan-beam geometry employing the entire beam transmitting the examined layer. For that, a flexible, arc-bent x-ray film was suggested as a 2D detector to record the sinogram directly. Such a geometry was only realized much later in commercial 3rd generation CT scanners, reducing the scanning time significantly. The proposed reconstruction procedure is based on an analog computing circuit coupled with a mechanical data readout and testifies an incredible ability to solve complex problems with limited technological resources. Even if the estimated reconstruction time of 5 min for a 100x100 matrix appears long, it was clearly way ahead of its time. Note, that Hounsfield used a powerful, state-of-the-art mainframe computer at EMI to reconstruct images from his first prototype scanner and it took about 2.5 hours to reconstruct a matrix of 80x80 pixels \cite{nref4}. The progress in computed tomography went incredibly fast after Hounsfield’s first demonstration of clinical application. In October 1971, the first patient was scanned in South London with Hounsfield’s first clinical prototype and in 1976 (about 5 years later), there were already 17 companies on the market offering commercial 3rd generation CT scanners \cite{nref4}. In 1979 Cormack and Hounsfield were awarded the Nobel Prize in Physiology or Medicine for their introduction of CT.

Reviewing the early works on CT by Tetelbaum, Korenblum and Tyutin, they seem to be the first to have published the idea to reconstruct the linear attenuation coefficient in a cross-section of an object from projectional x-ray data. In fact, they even solved the special case of the fan-beam geometry and proposed a respective apparatus and reconstruction method. They even envisioned time-resolved CT as well as employing different x-ray energies for the acquisition of "colored" images with additional diagnostic value \cite{nref6}. Most likely they were also the first to start an experimental realization before 1958. However, their efforts seem to have ended probably with the death of Tetelbaum in November 1958, since he must have been the engineering mind behind the hardware implementation and the computational system. 
Unfortunately, these early papers remained completely unnoticed in Western literature until 1983, when a short letter to the editor by Barrett et al. was published \cite{nref5}. Still those works remained widely unknown until today due to limited accessibility and the language barrier. Since the authors are hardly known related to computed tomography short biographies will be given in the following.

\textbf{Boris I. Korenblum} (orig. \foreignlanguage{russian}{Борис Исаакович Коренблюм}, 12 August 1923, Odessa – 15 December 2011, Slingerlands, New York) first started an education to become a violinist at the School of Stolyarsky in Odessa. However, after winning a young mathematician’s competition, he was given the opportunity to move to Kiev and pursue his study of mathematics. When World War II started the family left Kiev and – unlike the majority of the Jews in the capital – escaped the ani-Semitic massacres by Nazi Germany. Korenblum, who was not yet 17 at that time, volunteered for the army and served as a scout and interpreter due to his excellent German language skills. After the war he continued with his studies at the Institute of Mathematics of the National Academy of Sciences of Ukraine and received his Candidate of Sciences degree (equivalent to PhD) in 1947. He continued to work at the Institute, however, lost his position at the height of an anti-Semitic campaign in 1952 and considered himself lucky to secure a position at a Construction Engineering Institute in Kiev where he worked until 1973. In 1956 he received his Russian Doctorate of Sciences from Moscow State University. In 1973 he immigrated to Israel and held a professorship of Mathematics at the University of Tel-Aviv from 1974 to 1977. After that Korenblum went to the US for a visiting position as a member of the School of Mathematics at the Institute of Advanced Study in Princeton and assumed a professorship at the University at Albany, SUNY from 1977 until his retirement as professor emeritus in 2009. Korenblum mainly researched classical harmonic analysis, functional analysis, Banach algebras (particularly Bergman spaces) and complex analysis. He died of natural causes at his home in Slingerlands, NY on December 15, 2011 at the age of 88.   

\begin{figure}
  \centering
  \includegraphics[width=12cm]{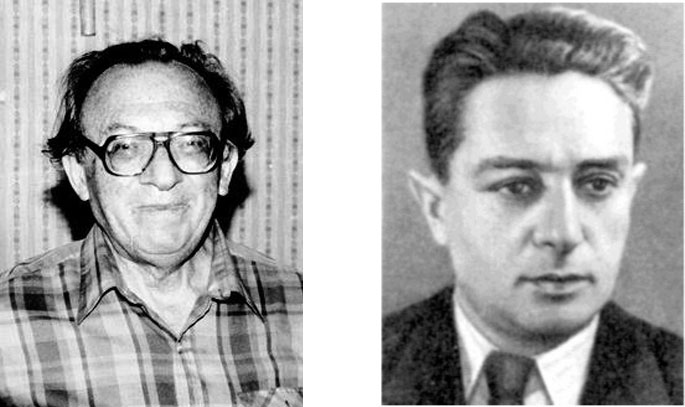}
  \caption{Boris I. Korenblum (left) and Semyon I. Tetelbaum (right)}
  \label{fig:fig3}
\end{figure}

\textbf{Semyon I. Tetelbaum} (orig. \foreignlanguage{russian}{ Семён Исаакович Тетельбаум}, July 7, 1910, Kiev - November 24, 1958, Kiev) studied at the Kiev Polytechnic Institute (KPI) from 1928-1932 and worked there as an electrical design engineer performing pioneering contributions to television and radio. In 1932 he designed and built a television system enabling first experiments on television in Ukraine. In the later Thirties he carried out research on radio transmission and supervised the development and construction of radio stations in Kiev and Odessa. He defended his doctoral thesis in 1939 and became professor and head of the department of "Receiving and Transmitting Technology" in 1940 at KPI. During Word War II he was Dean of the Radio Engineering Department of the Central Asian Industrial Institute and worked on radar technology and high frequency-induced melting of metal alloys. From 1946 he headed the high-frequency current laboratory at the Institute of Electrical Engineering of the Academy of Sciences of Ukraine and became an elected member of the latter in 1948. He served as the first managing editor of the Soviet Union-wide scientific and technical journal "Proceedings of the Higher Educational Institutions in the Field of Radio Technology" and taught various courses related to electrical engineering at the Kiev Polytechnic Institute. He also investigated problems of efficient wireless energy transmission, sonar, remote control, automatization and electronics in medicine. The theoretical and technical development of computed tomography with x-rays was among his last projects. He died unexpectedly on November 24 in 1958 at the age of 48 in Kiev.

I was not able to find biographical data on the last author, \textbf{A. A. Tyutin} (orig. \foreignlanguage{russian}{А. А. Тютин}) and would be thankful for any hint or remark.

\section{Acknowledgements}

I would like to thank Fabio De Marco for bringing the work to my attention and further discussion and research on original sources as well as feedback on the translation. I want to acknowledge the assistance of the staff members of the Scientific and Technical Library of Igor Sikorsky Kiev Polytechnic Institute with providing copies of the original work and some of the references. I am grateful to Daniel Korenblum for reviewing my translation and Marina Korenblum for providing biographical information on Boris Korenblum.  

\section{Appendix A}

Scans of the original work in Russian language are provided below. 

\begin{figure}
  \centering
  \includegraphics[width=15cm]{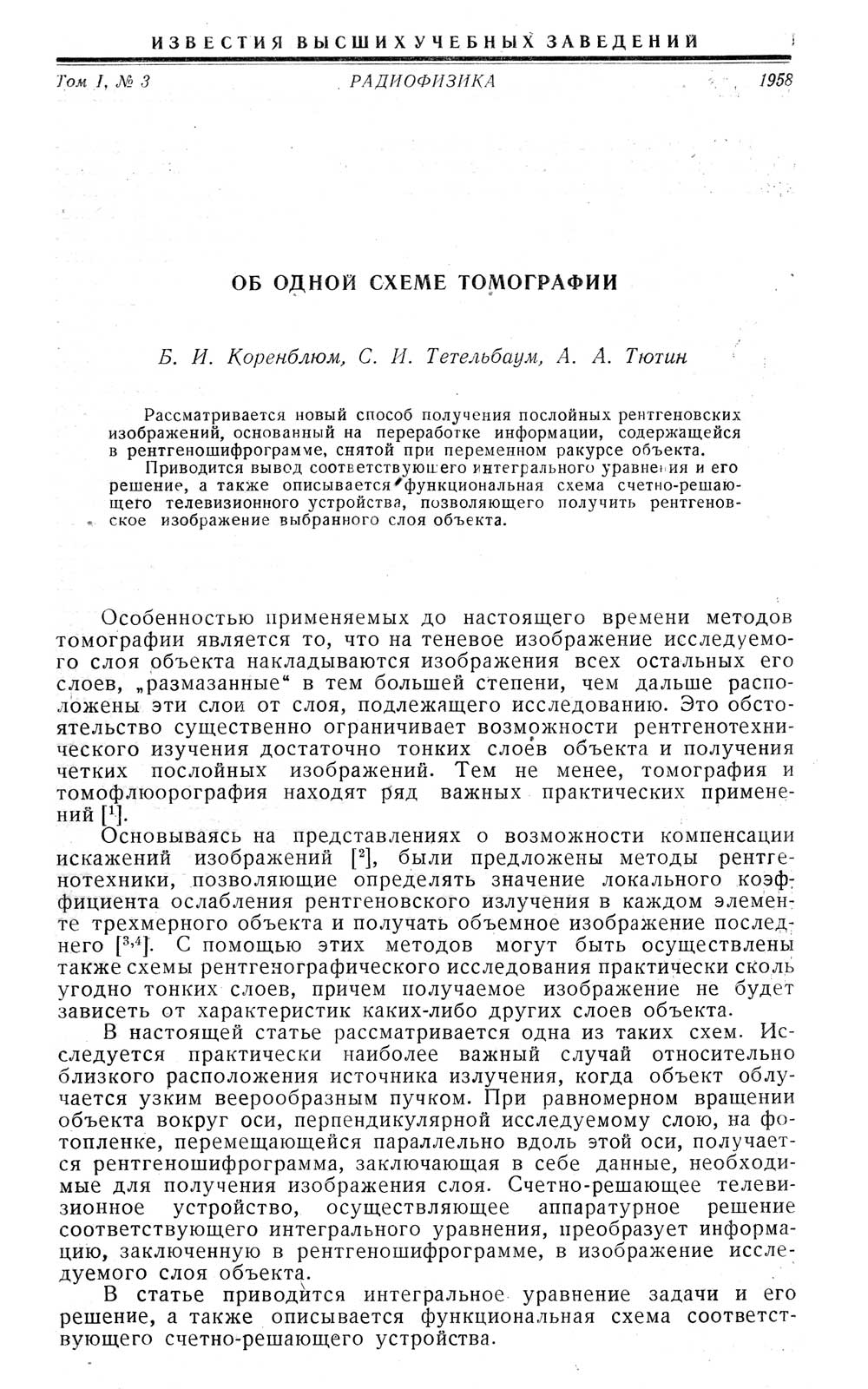}
\end{figure}

\begin{figure}
  \centering
  \includegraphics[width=15cm]{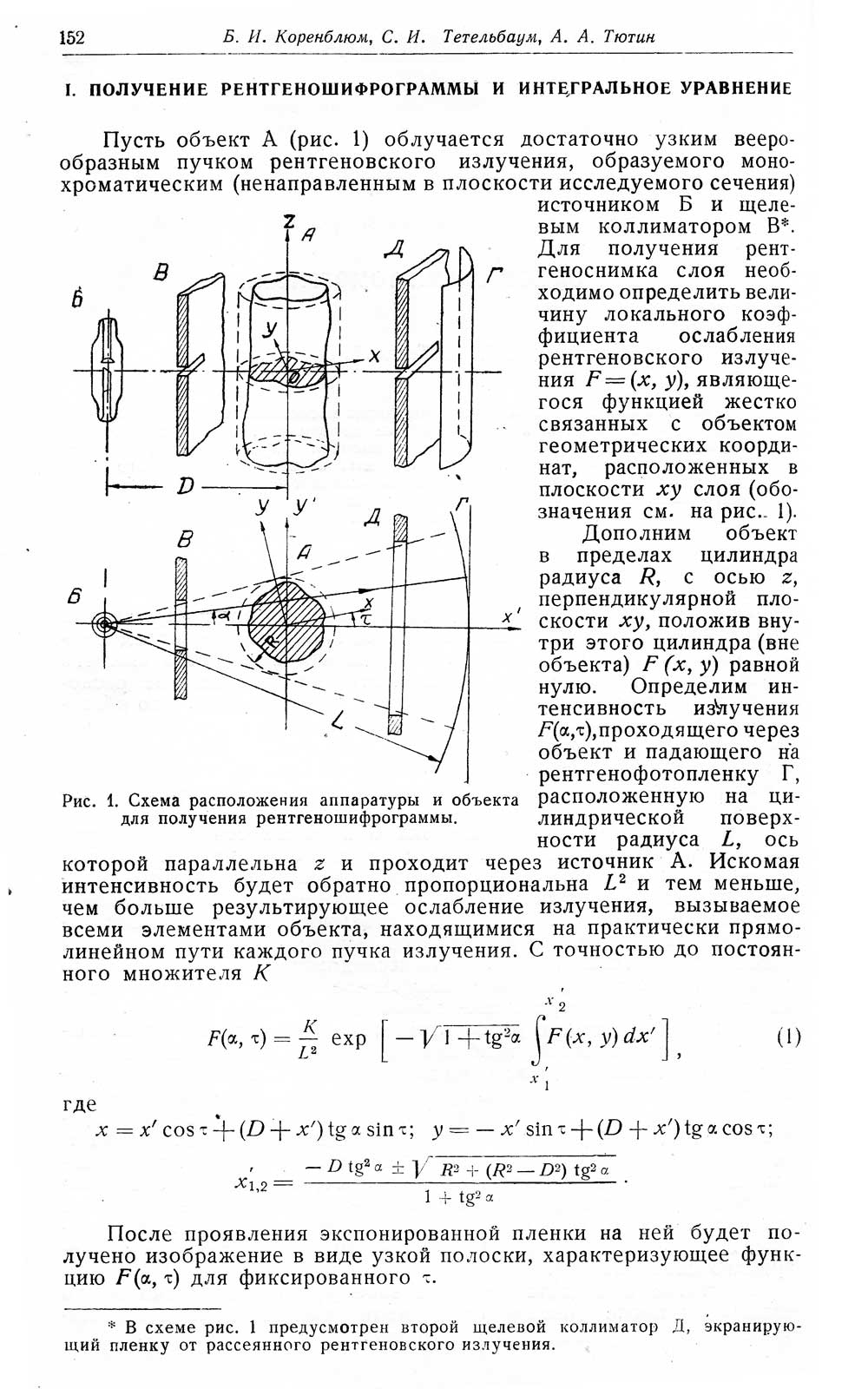}
\end{figure}

\begin{figure}
  \centering
  \includegraphics[width=15cm]{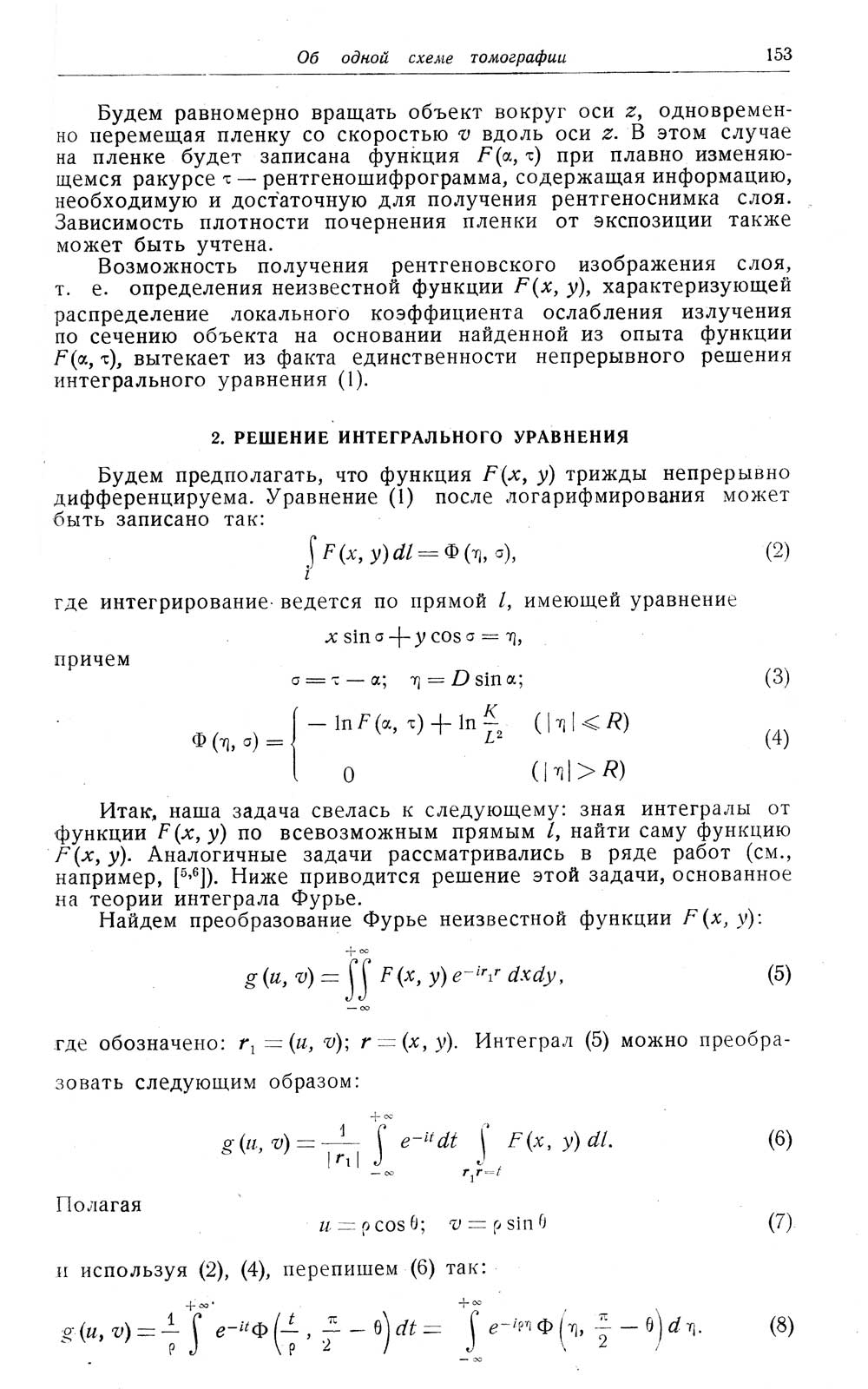}
\end{figure}

\begin{figure}
  \centering
  \includegraphics[width=15cm]{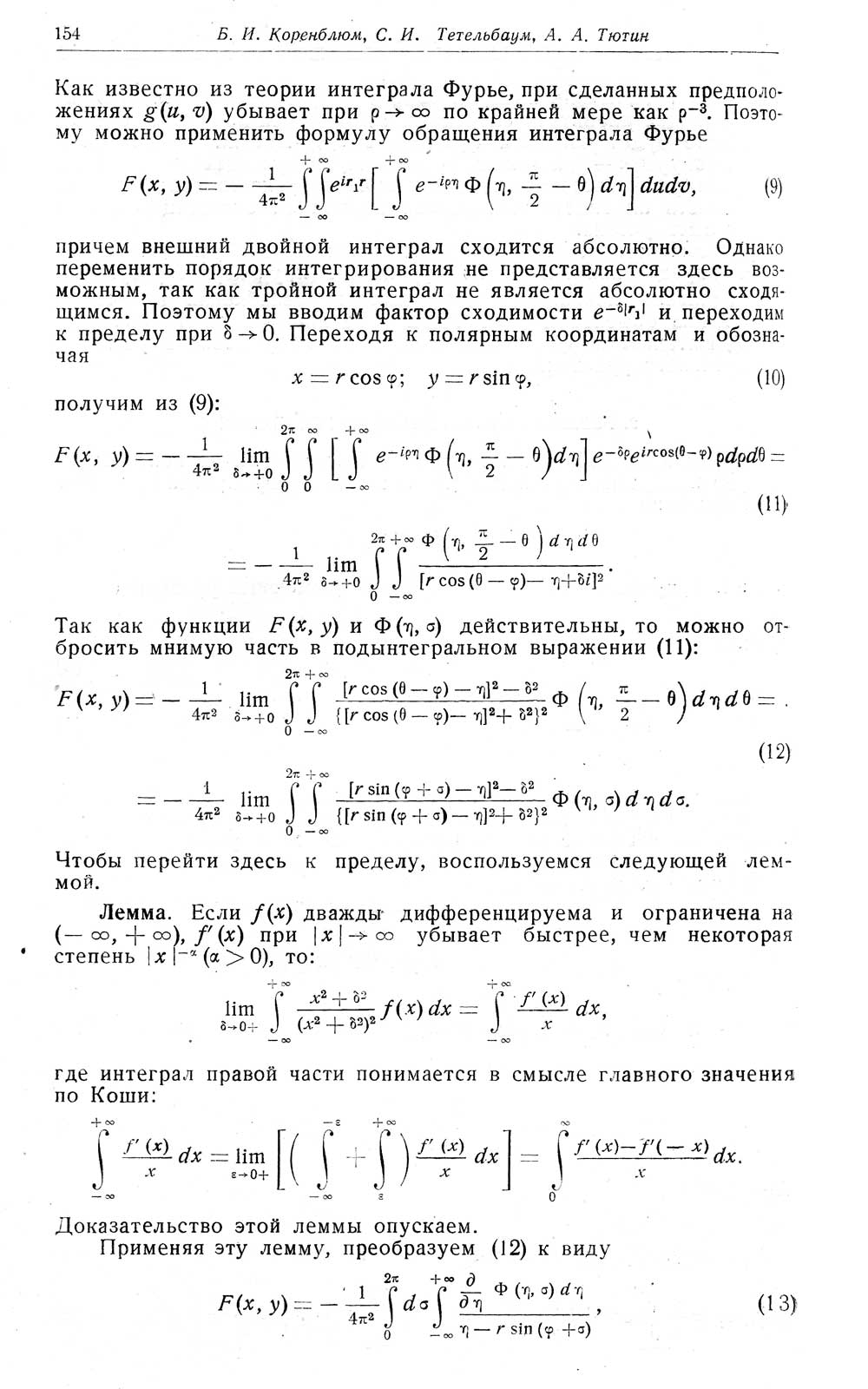}
\end{figure}

\begin{figure}
  \centering
  \includegraphics[width=15cm]{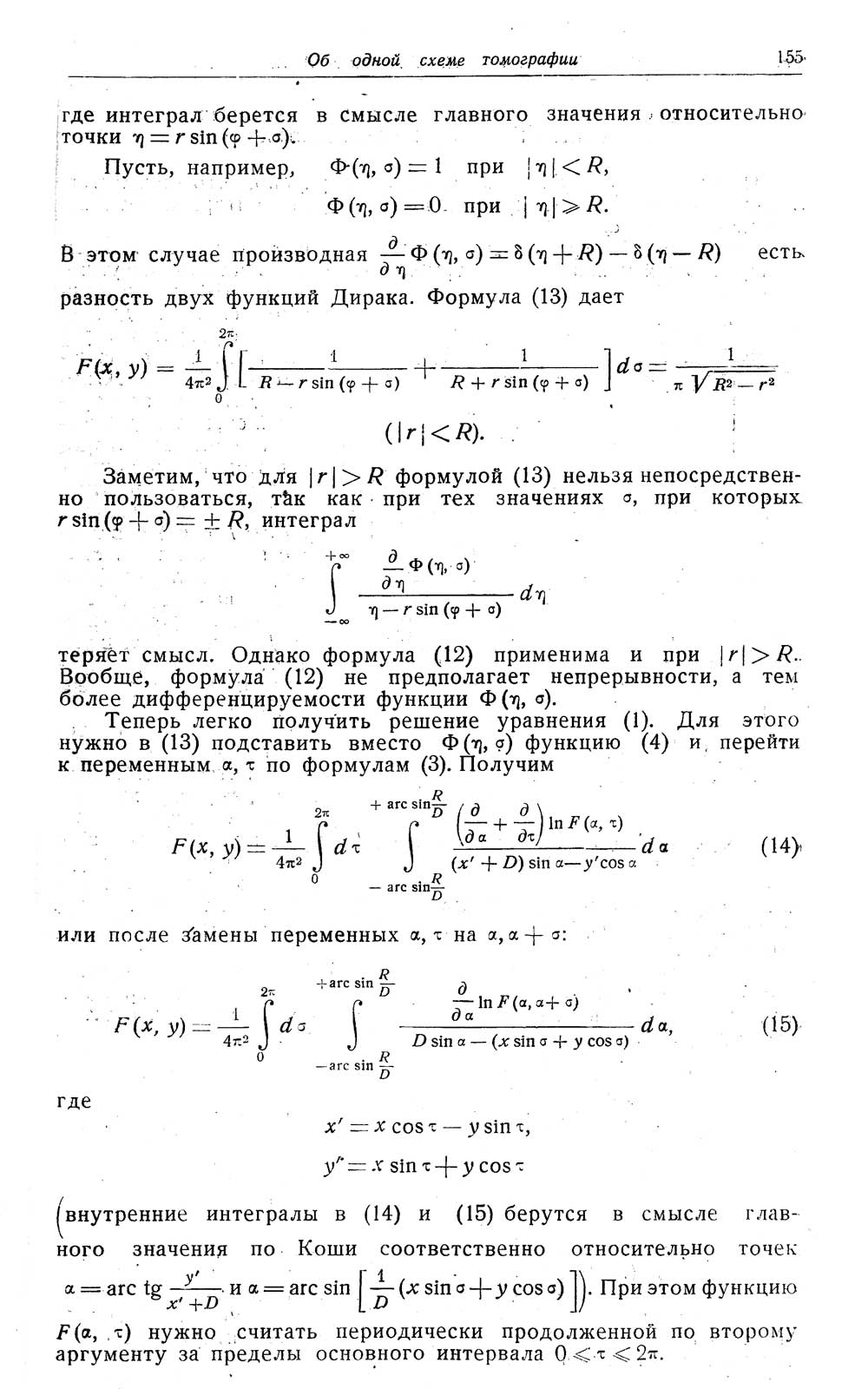}
\end{figure}

\begin{figure}
  \centering
  \includegraphics[width=15cm]{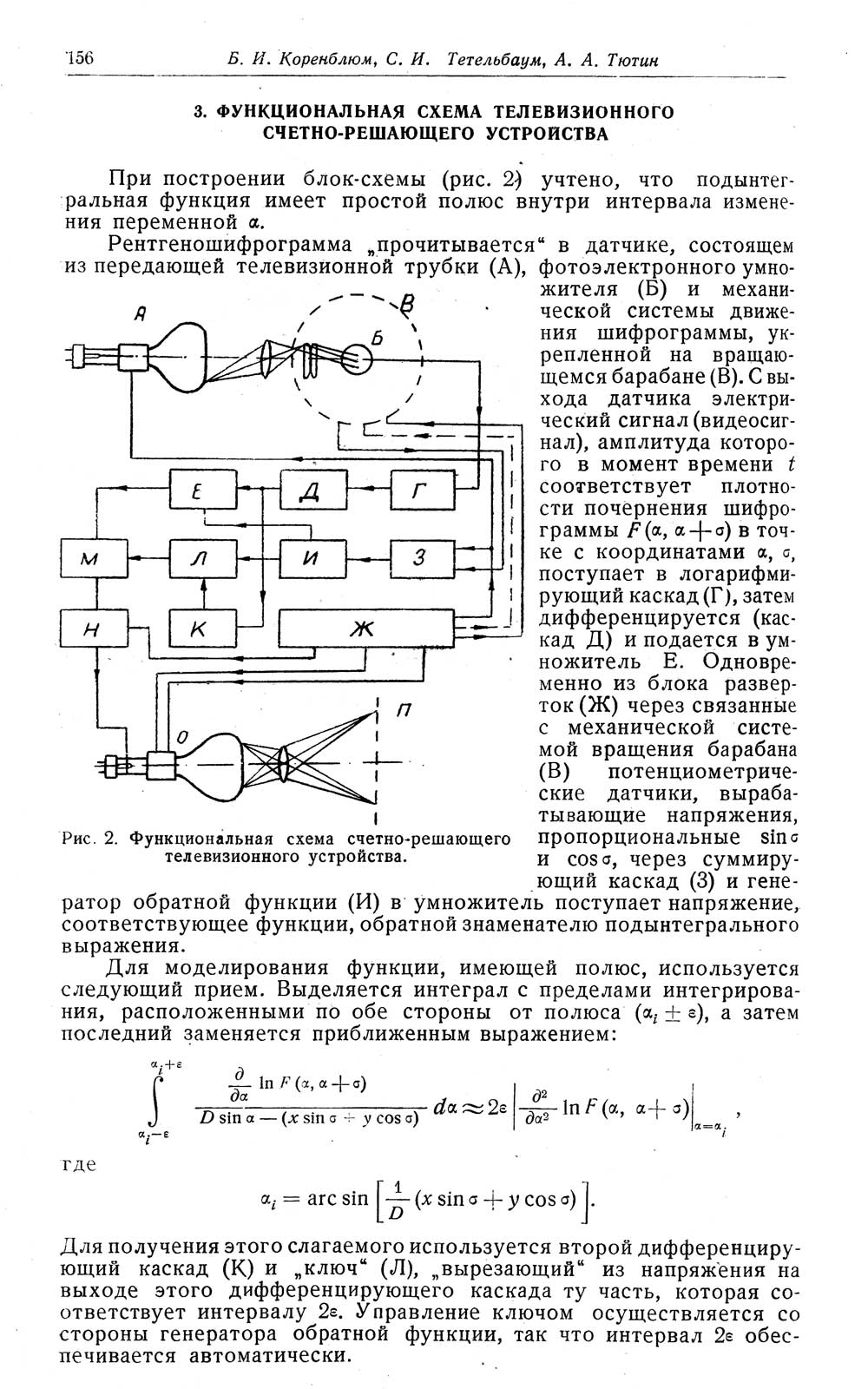}
\end{figure}

\begin{figure}
  \centering
  \includegraphics[width=15cm]{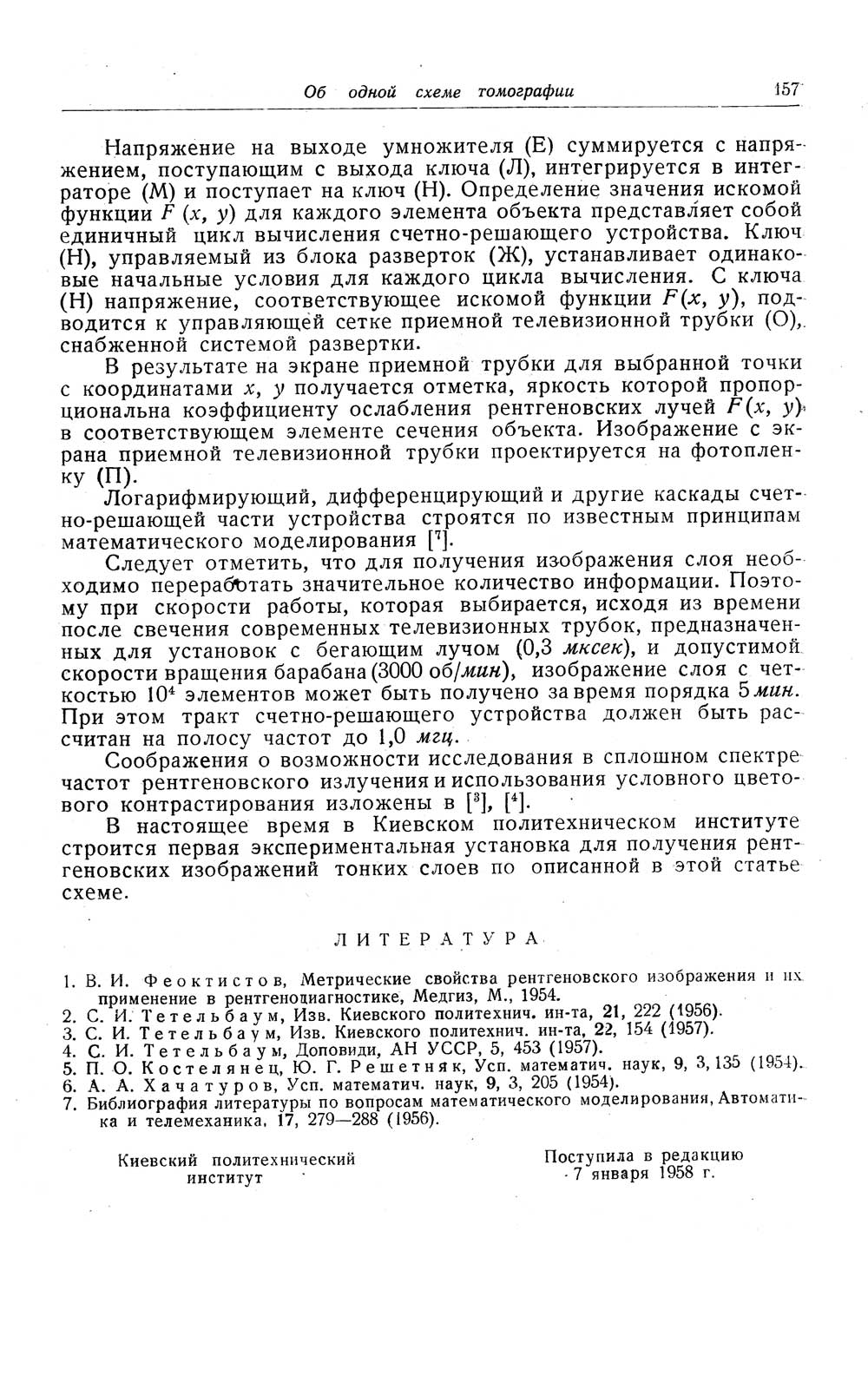}
\end{figure}

\bibliographystyle{unsrt}  


\end{document}